\documentclass{aa}
\usepackage{graphicx}
\usepackage{txfonts}

\def\lsim{\;\raise0.3ex\hbox{$<$\kern-0.75em\raise-1.1ex\hbox{$\sim$}}\;}
\def\gsim{\;\raise0.3ex\hbox{$>$\kern-0.75em\raise-1.1ex\hbox{$\sim$}}\;}
\def\gr{$\gamma$-ray }
\def\gc{$\gamma$-Cyg }
\def\etal{et al. }

\def\beq{\begin{equation}}
\def\enq{\end{equation}}
\def\begar{\begin{eqnarray}}
\def\endar{\end{eqnarray}}
\def\mathnew{\mathsurround=0pt}
\def\simov#1#2{\lower .5pt\vbox{\baselineskip0pt \lineskip-.5pt
        \ialign{$\mathnew#1\hfil##\hfil$\crcr#2\crcr\sim\crcr}}}

\def\cmc{\rm ~cm^{-3}}
\def\cms{\rm ~cm^{-2}}
\def\kms{\rm ~km~s^{-1}}

\def\enf{\rm ~erg~cm^{-2}~s^{-1}}
\def\fl{\rm ~cm^{-2}~s^{-1}}
\def\phfl{\rm ~ph~cm^{-2}~s^{-1}~keV^{-1}}
\def\lfl{\rm ~ph~cm^{-2}~s^{-1}}
\def\nh{N$_H$}
\def \rchisq {$\chi_{\nu} ^{2}$}
\def\approxgt{\mathrel{\hbox{\rlap{\lower.55ex \hbox {$\sim$}}
        \kern-.3em \raise.4ex \hbox{$>$}}}}
\def\approxlt{\mathrel{\hbox{\rlap{\lower.55ex \hbox {$\sim$}}
        \kern-.3em \raise.4ex \hbox{$<$}}}}
\begin{document}
\title{\bf
Hard X-ray Emission Clumps in the $\gamma$-Cygni Supernova Remnant:
an INTEGRAL-ISGRI View}

   \author{
             A.M.Bykov
     \inst{1},
     A.M.Krassilchtchikov
     \inst{1},
      Yu.A.Uvarov
     \inst{1},
     H.Bloemen
     \inst{2},
      R.A.Chevalier
     \inst{3},
      M.Yu.Gustov
     \inst{1},
      W.Hermsen
     \inst{2},
     F.Lebrun
     \inst{4},
   T.A.Lozinskaya
     \inst{5},
     G.Rauw
     \inst{6},
    T.V.Smirnova
     \inst{7},
      S.J.Sturner
     \inst{8},
      J.-P.\,Swings
      \inst{6},
      R.Terrier
     \inst{4},
      I.N.Toptygin
     \inst{1}
          }

   \offprints{A.M.Bykov~(byk@astro.ioffe.ru)}

   \institute{A.F.Ioffe Institute for Physics and Technology,
              26 Polytechnicheskaia, 194021, St.Petersburg, Russia 
       \and
         SRON National Institute for Space Research, Sorbonnelaan 2,
         3584 CA Utrecht, The Netherlands 
       \and
         Department of Astronomy, University of Virginia,
         P.O. Box 3818, Charlottesville,  VA 22903, USA 
       \and
         CEA-Saclay, DSM/DAPNIA/Service d'Astrophysique,
         91191 Gif-sur-Yvette Cedex, France 
 \and
     Sternberg Astronomical Institute, Moscow State University,
     13 Universitetskij, 119899, Moscow, Russia 
     \and
      Institut d'Astrophysique et de G\'eophysique, Universit\'e de Li\`ege,
       All\'ee du 6 Ao\^ut 17, B\^at B5c, 4000 Li\`ege, Belgium 
     \and
     Astro Space Center of the Lebedev Physics Institute,
     84/32 Profsoyuznaia, 117810, Moscow, Russia 
     \and
     NASA Goddard Space Flight Center, Code 661, Greenbelt, MD 20771, USA 
             }

   \date{Received September 22, 2004; accepted October 9, 2004}

\abstract{ Spatially resolved images of the galactic supernova
remnant G78.2+2.1 ($\gamma$-Cygni) in hard X-ray energy bands from
25 keV to 120 keV are obtained with the {\it IBIS-ISGRI} imager
aboard the International Gamma-Ray Astrophysics Laboratory {\it
INTEGRAL}. The images are dominated by localized clumps of about
ten arcmin in size. The flux of the most prominent North-Western
(NW) clump is (1.7$\pm$0.4)$\times$10$^{-11} \enf$ in the 25-40
keV band. The observed X-ray fluxes are in agreement with
extrapolations of soft X-ray imaging observations of \gc by {\it
ASCA GIS} and spatially unresolved {\it RXTE PCA} data. The
positions of the hard X-ray clumps correlate with bright patches
of optical line emission, possibly indicating the presence of
radiative shock waves in a shocked cloud. The observed spatial
structure and spectra are consistent with model predictions of
hard X-ray emission from nonthermal electrons accelerated by a
radiative shock in a supernova interacting with an interstellar
cloud, but the powerful stellar wind  of the O9V star HD 193322  is a
plausible candidate for the NW source as well.

}

\authorrunning{A.M.Bykov et al.}
\titlerunning{Hard X-ray Emission Clumps in $\gamma$-Cygni}
\maketitle

\section{\bf Introduction}

The supernova remnant (SNR) G78.2+2.1 ($\gamma$-Cygni) is a degree
size extended source which has been imaged in radio
waves to $\gamma$-rays.   
The SNR is located in the complex Cygnus X region of  massive gas
and dust complexes and close to the most powerful Cyg OB2
association. Radio observations by Higgs \etal (1977) established
a shell-like structure of the remnant, suggested also by
Lozinskaya (1977). Multi-frequency radio observations by
Zhang \etal (1997) revealed a patchy structure of the radio
spectral-index distribution. The integrated radio spectral index
$\alpha$ = 0.54 $\pm$ 0.02, with variations of $\pm$ 0.15
within the remnant. Prominent radio brightness
enhancements are present in the South-East and the North-West of
\gc (e.g. Zhang \etal 1997). A distance to \gc of $\sim$ 1.5 kpc
was estimated by Landecker \etal (1980) from radio HI
observations.

Optical images of \gc in H$_{\alpha}$+[NII], [SII] and [OIII]
filters, recently presented by Mavromatakis (2003), clearly show a
patchy structure with bright spots of some ten arcminutes scale
and line emission fluxes of a few times $\,10^{-15}$ erg cm$^{-2}$
s$^{-1}$ arcsec$^{-2}$ for H$_{\alpha}$ filters and a few times
$\,10^{-16}$ erg cm$^{-2}$ s$^{-1}$ arcsec$^{-2}$ for the [OIII]
filter.

Archival {\it ROSAT} and {\it ASCA} observations of G78.2+2.1 were
analyzed by Lozinskaya \etal (2000). They pointed out a complex
structure of the remnant with {\it ROSAT} emission extending well
beyond the apparent SNR radio shell, possibly indicating expansion
into a progenitor star wind cavity. With archival {\it ASCA}
observations, Uchiyama \etal (2002) found X-ray emission above 4
keV to be dominated by several localized clumps, mostly in the
Northern part of the remnant. The 4-10 keV emission of the  clumps
has a hard photon index of 0.8--1.5.

A high-energy \gr source, 2CG 078+2, was discovered in the field
of \gc with the {\it COS B} satellite (e.g. Swanenburg \etal
1981). {\it CGRO-EGRET} confirmed this source (2EG J2020+4026/3EG
J2020+4017), which is one of the brightest steady-state
unidentified sources in the {\it EGRET} catalogue with a flux of
$\sim$ 1.2 $\times 10^{-6}\fl$ above 100 MeV (Sturner and Dermer
1995; Esposito \etal 1996). {\it Whipple} \gr telescope
observations (e.g. Buckley \etal 1998) established an upper limit
of 2.2 $\times 10^{-11}\fl$ for the flux above 300 GeV, indicating
a break in the high-energy emission spectrum above a few GeV. The
\gr emission might be produced by a pulsar, but can be attributed
to interactions of accelerated energetic particles with ambient
matter and radiation as well (e.g. Sturner \etal 1997; Gaisser \etal 1998;
Bykov \etal 2000). Solving this problem of the origin
of the \gr emission requires the analysis of multiwavelength
imaging observations. Hard X-ray imaging provides a crucial tool
to distinguish between leptonic and hadronic contributions to the
\gr emission. We present below the first hard X-ray imaging
observation of \gc.

\begin{figure}
\includegraphics[width=0.5\textwidth]{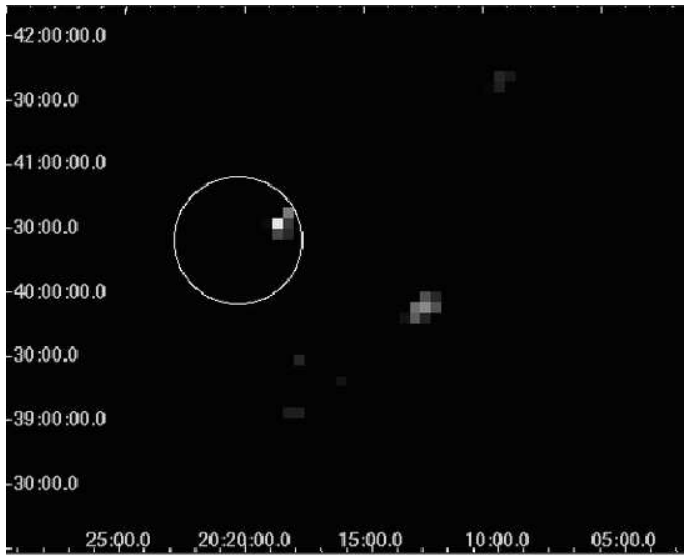}
\caption{The {\it ISGRI} 25--40 keV broad view of the \gc region
(linearly scaled S/N map of the sources above  2$\sigma$).
The solid circle is the same as in Fig.~\ref{fig:A_R_I} .
}
\label{fig:broad_view}
\end{figure}

\section{Observations and Data Analysis}

\subsection{INTEGRAL ISGRI observations}

The SNR G78.2+2.1 was observed with the {\it ISGRI\,} imager (Lebrun
\etal 2003) aboard {\it INTEGRAL} (Winkler \etal 2003). We
combined data from 21 Core Program and calibration observations
(revolutions 12-82: Nov 18, 2002 --- Jun 15, 2003) and added 28
pointings from AO1 (rev. 80, 11-12 Jun 2003) and AO2 (rev. 191,
8-9 May 2004) observations [Obs. Ids 0129700 and 0229700:
amalgamated observations with G. Rauw as PI].

The data obtained from {\it ISGRI\,} have been reduced with the
standard off-line scientific analysis software developed at the
INTEGRAL Science Data Center (the OSA 4.0 package). The standard
good time selection criteria were applied; only science windows
(SCWs) with more than 100~s of good time were considered. Wide
energy bands were used (25-40 keV, 40-80 keV, and 80-120 keV) in
order to be able to detect sources at a few mCrab level. In order
to improve on source detection (at the cost of source-localization
accuracy), pixel spreading was switched off. We analysed 360 good
ks of FCFOV (fully coded field of view) observations and 1,060
good ks of PCFOV (partially coded field of view) observations,
although only FCFOV data were used for the flux estimates
presented here. Source reconstruction with FCFOV data is more
secure in the Cygnus region with its bright point sources and
complex diffuse background emission. The angular resolution (FWHM)
of {\it ISGRI} is about 12\arcmin\, and the images were sampled in
$\sim$ 5\arcmin\, pixels (Lebrun \etal 2003; Goldwurm \etal 2003).

\begin{figure*}
\includegraphics[width=0.33\textwidth]{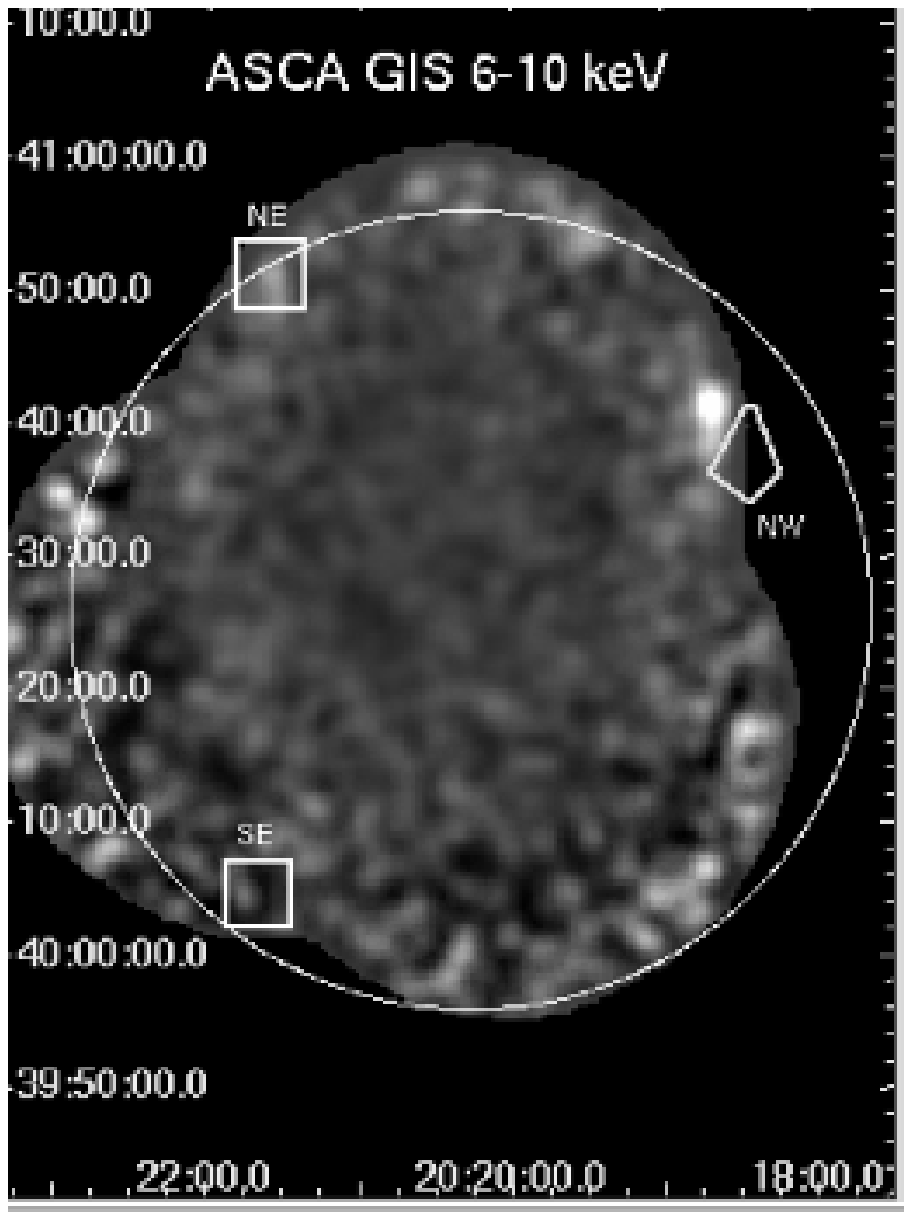}
\includegraphics[width=0.33\textwidth]{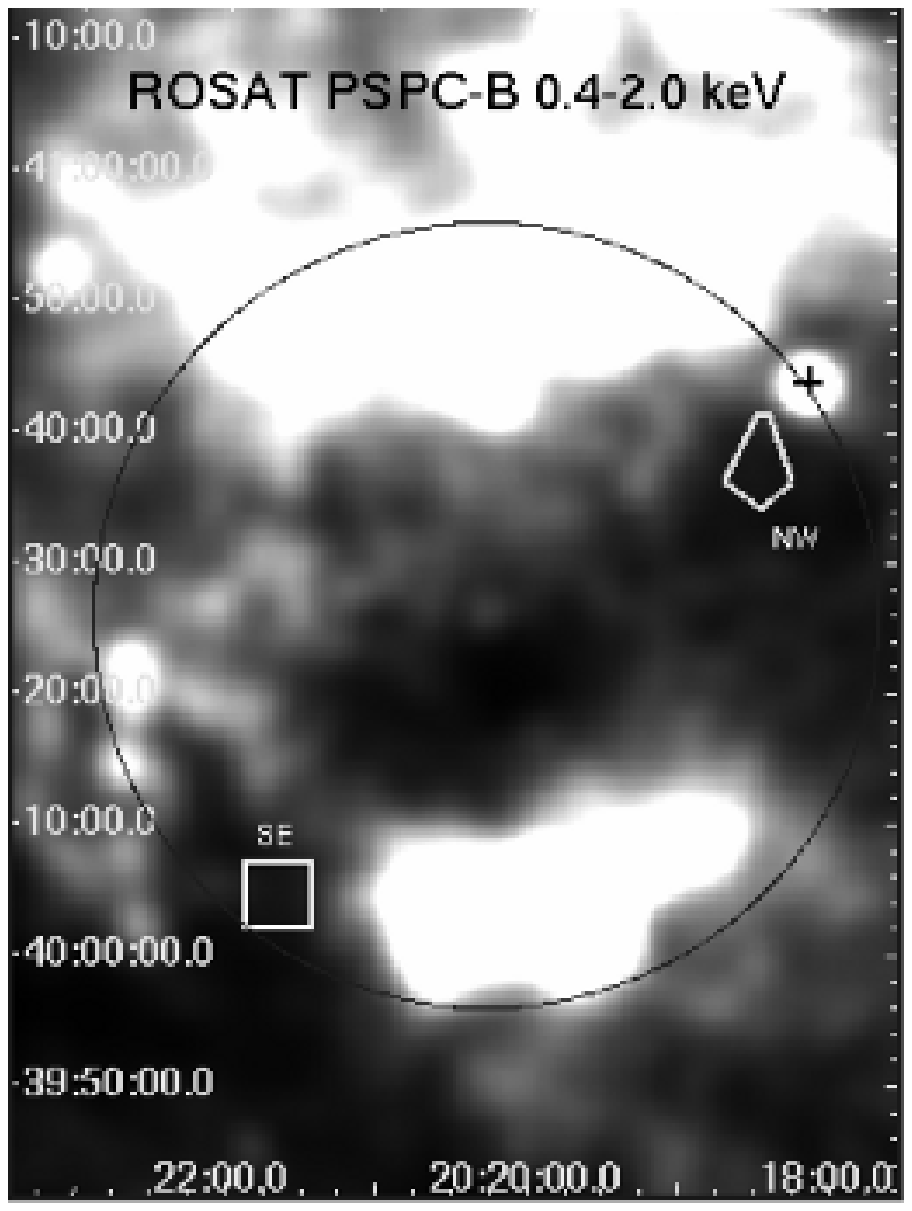}
\includegraphics[width=0.33\textwidth]{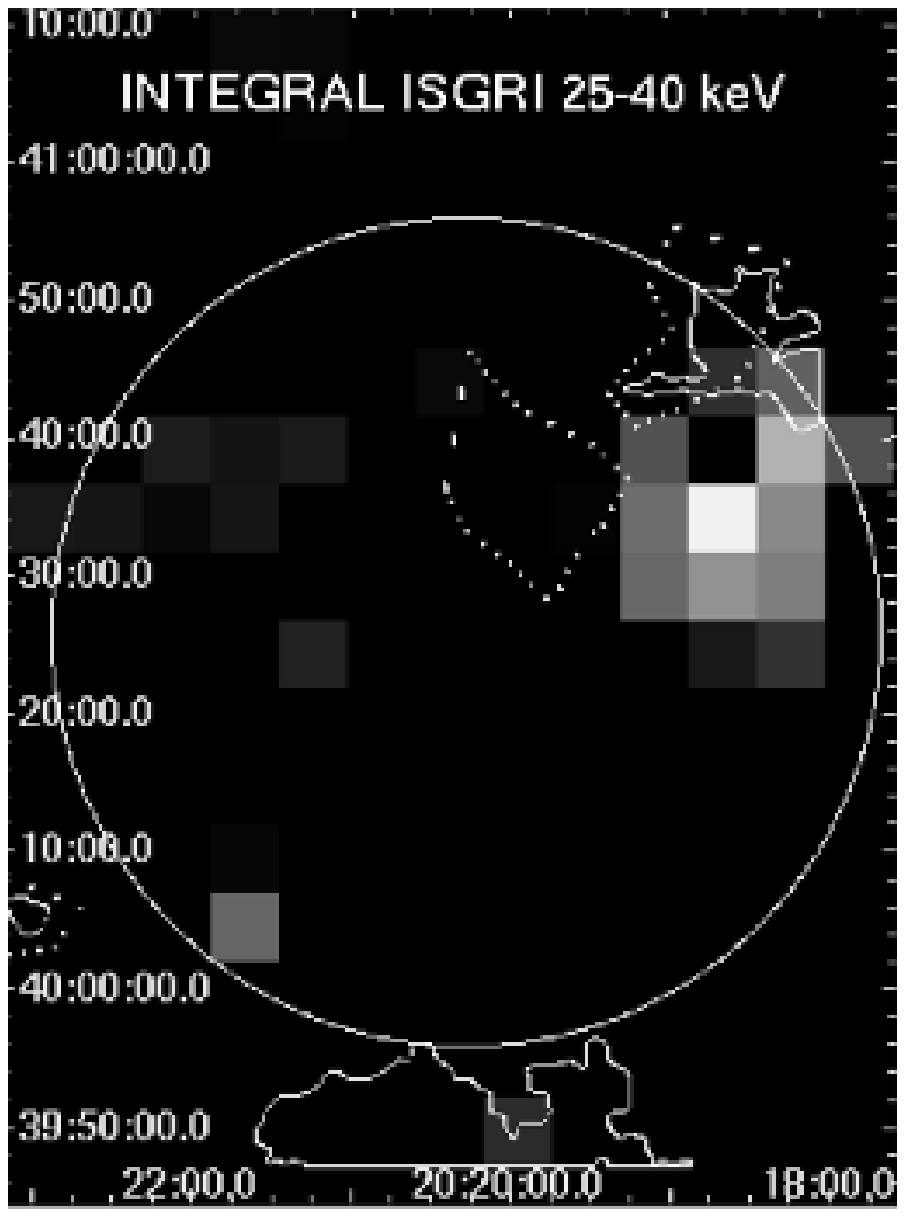}
\caption{{\it Left}: {\it ASCA GIS} 6-10 keV grayscale map with an
{\it INTEGRAL ISGRI} 25-40 keV 4.2 $\sigma$ contour in the NW
area. {\it Centre}: {\it ROSAT PSPC-B} 0.4-2.0 keV grayscale map
with the same {\it INTEGRAL ISGRI} 25-40 keV 4.2 $\sigma$ contour.
{\it Right}: {\it INTEGRAL ISGRI} 25-40 keV significance map
(linear scale) with H$_{\alpha}$+[NII] 6560 \AA\,
3$\times$10$^{-15}$ erg cm$^{-2}$ s$^{-1}$ arcsec$^{-1}$ contours
(dotted line) and [OIII] 5010 \AA\, 3$\times$10$^{-16}$ erg
cm$^{-2}$ s$^{-1}$ arcsec$^{-1}$ contours (solid line) [optical
data are from Mavromatakis (2003)]. The hard emission regions NW,
NE, and SE are marked in the first two panels. The NE clump is
seen with {\it ISGRI} only in the 40-80 keV band. The {\it RXTE
PCA} field of view is indicated by the large solid circle. The
black cross on the {\it ROSAT} map marks the position of the O9V
star HD 193322 (20:18:07,+40:43:55). All maps are made for J2000.}


\label{fig:A_R_I}
\end{figure*}


Source significances were derived
from a statistical distribution of count-rate values for a sample of
about 100$\times$100 pixels in {\it ISGRI}\, FCFOV images around
\gc (using the mean and the standard deviation of the distribution).

At present, the most reliable {\it ISGRI} flux estimates can be
obtained by cross calibration with the Crab, i.e. by comparing the
source and Crab count rates in the same energy bands, accounting
for different [model] spectral distributions.  The systematic
uncertainties of this procedure are estimated to be about 25\% ,
mainly due to the possibly extended nature ($\gsim$ 10\arcmin) of
the detected sources (cf. Lubinski 2004 where calibrations of
point sources are discussed).

In Figs.~\ref{fig:broad_view},~\ref{fig:A_R_I} we show X-ray images of G78.2+2.1 obtained
with {\it INTEGRAL ISGRI}, {\it ASCA}, and {\it ROSAT}.
The {\it ISGRI} images in
our three energy bands are dominated by a few localized clumps of
emission. The flux estimates of the North-Western (NW) clump are
$(1.7\pm0.4)\times 10^{-11} \enf$ in the 25--40 keV band,
$(1.2\pm0.8)\times 10^{-11} \enf$ in the 40-80 keV band, and
$(2.5\pm1.2)\times 10^{-11} \enf$ in the 80--120 keV band. For the
SE clump we find $(1.2\pm0.5)\times 10^{-11} \enf$ (25-40 keV),
$(1.5\pm0.8) \times 10^{-11} \enf$ (40-80 keV), and
$(3.0\pm1.2)\times 10^{-11} \enf$ (80-120 keV). The NE clump that
coincides with the ASCA hard source C1 (Uchiyama \etal 2002) is
seen only in the 40-80 keV band with a flux of $(1.7\pm0.7)\times
10^{-11} \enf$. A longer exposure with {\it ISGRI} is required to
detect (or reject) this NE-clump emission; the weak detection
obtained so far might be related to the extended nature of the ASCA
C1 source.

\subsection{RXTE PCA and ASCA GIS observations}

Hard X-ray emission from G78.2+2.1 can be also constrained from
non-imaging {\it RXTE} observations. We used public archive {\it
RXTE} 61.4 ksec observations of \gc\ performed on 5-10 Apr 1997.
The {\it RXTE PCA} data were reduced with the standard HEASARC
FTOOLS 5.3
software.
Only Standard-2 mode data were used.
The PCA background was accounted for with the latest
faint source background model (cmfaintl7\_eMv20031123). The
standard good time selection procedure was applied. The {\it RXTE
PCA} field of view covers a substantial part of \gc\ including the
hard clumps seen with {\it INTEGRAL ISGRI}. We found the X-ray
emission of \gc\ to extend beyond $\sim$ 20 keV and fitted the
{\it RXTE PCA} 5-15 keV data simultaneously with archival {\it
ASCA GIS2} 3-7 keV data for the total remnant emission with a
broken power-law with a Lorentz line at 6.2 keV. Here we used the
same set of {\it ASCA GIS} data that was analyzed by Uchiyama
\etal (2002).

The total remnant emission (3-15 keV) was fitted by a power-law
with photon indexes $\alpha_1 = 2.0 \pm 0.4$ below the break
energy $E_b = 11.1 \pm 1.2$ keV and $\alpha_2 = 1.2 \pm 0.4$ for
$E_b < E < 15$ keV (the reduced \rchisq = 0.91 at $\nu$=85
d.o.f.). The power-law normalization was (8.1 $\pm$ 0.56) $\times$
10$^{-3} \phfl$ at 1 keV.  We found a correction factor (0.72
$\pm$ 0.19) for the {\it RXTE PCA} flux normalization considered
as a free parameter with a fixed {\it ASCA GIS2} flux
normalization. The line at 6.2 $\pm$ 0.04 keV has a width of 1.0
$\pm$ 0.2 keV (equivalent width is 1.1 $\pm$ 0.5 keV) and
normalization (2.5 $\pm$ 0.8) $\times$ 10$^{-4} \lfl$. Note here
that a similar signature of a 6.4 keV line was found by Pannuti
\etal (2003) in the {\it RXTE} analysis of another extended SNR
G347.3-0.5. While the low energy branch of the fit is a
combination of thermal and nonthermal components, the high energy
emission above the break is nonthermal.  Using the joint {\it ASCA
GIS2 - RXTE PCA} fit, we estimated the ranges of the extrapolated
hard X-ray fluxes of the total remnant emission in the 25-40 keV
band as (1.0 - 1.7)$\times 10^{-11} \enf$, while in the 80-120 keV
regime the range is (1.4 - 4.5)$\times 10^{-11} \enf$. Since the
hard X-ray {\it ISGRI} images are dominated by the NW, NE, and SE
clumps that are located at the boundary of the {\it RXTE PCA}
pointing, we have applied flux  corrections accounting for the
{\it RXTE PCA} collimator response at $\sim$ 30\arcmin\ from the
axis.

\begin{figure}
\includegraphics[width=0.5\textwidth,height=6.5cm,bb= 26 77 565 493,clip]{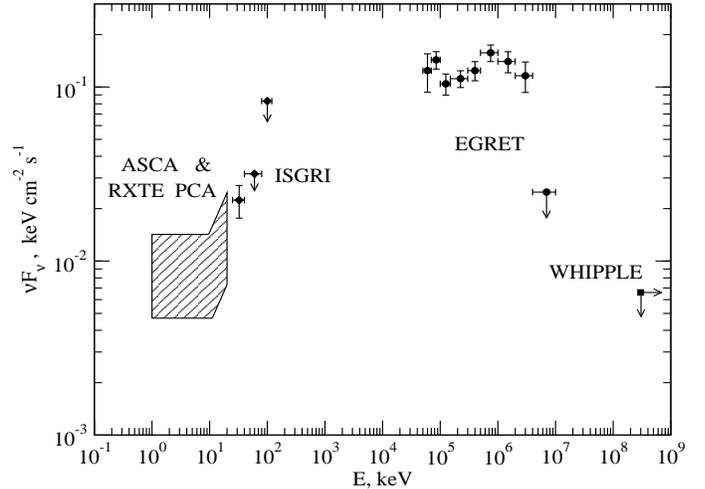}
\caption{The spectral energy distribution of the hard emission
from \gc. The {\it ISGRI} datapoints correspond to the NW clump
only.} \label{fig:spectra}
\end{figure}

The total flux  estimations are in agreement with the summed {\it
INTEGRAL ISGRI} fluxes  of the NW, NE, and SE clumps if a 25\%
systematic error in the calibration procedure is accounted for.
The hard-emission spectrum of \gc is presented in
Fig.~\ref{fig:spectra}. We show the total remnant {\it ASCA
GIS2 - RXTE PCA} flux, the fluxes  of the {\it ISGRI}\, NW clump and of {\it EGRET}\, 3EG J2020+4017 source, and the {\it Whipple} upper
limit.

\section{\bf Discussion}

The locations of the hard X-ray clumps detected by {\it INTEGRAL
ISGRI} correlate with dim regions of high \nh $\gsim 5\times
10^{21} \cms$ in the {\it ROSAT PSPC-B} 0.4-2.0 keV greyscale map
in Fig.~\ref{fig:A_R_I}. The clumps adjoin the patches of bright
optical H$_{\alpha}$+[NII], [SII] and [OIII] line emission (of
$\gsim$ 10\arcmin\ scale size) observed by Mavromatakis (2003).
The line emission may indicate an interaction of the SNR with a
nearby cloud and the presence of a radiative shock wave.

In evolved supernova remnants interacting with interstellar
clouds, such as \gc\ and IC~443, a highly inhomogeneous structure
consisting of a forward shock of a moderate Mach number, a cooling
layer, a dense radiative shell, and an interior region filled with
hot tenuous plasma is expected. A SNR evolving in
the inter-clump medium of a molecular cloud of density $n_a \sim
25$ H atoms cm$^{-3}$  becomes radiative at radii $\sim 6$ pc
(Chevalier 1999). A model of nonthermal electron acceleration and
interactions in such SNRs by Bykov \etal (2000) predicted that
these SNRs are efficient electron accelerators and sources of hard
X- and \gr emission.
A distinctive feature of the model is the presence of emitting
clumps with very flat hard X-ray spectra of photon indexes
$\alpha \sim 1$. The photon break energy and the maximum energy of
accelerated electrons depend on the ambient density. A radiative
shock propagating through the interclump medium could accelerate
radio-emitting electrons in the GeV regime, also producing hard X-ray
and \gr {\it EGRET}-regime emission. A slower MHD shock in a dense
molecular clump could produce bright hard X-ray emission without
prominent radio and \gr counterparts.  Bremsstrahlung radiation from
the nonthermal electrons has a hard X-ray photon spectrum. The
nonthermal X-ray flux  $F_x$ from the electrons accelerated by
a MHD shock of a velocity $v_{s7}$ (in 100 $\kms$) and of an
apparent angular size $\theta$ (in arcmin) can be estimated as a
fraction of the shock ram pressure:
\begin{equation}
F_x \approx 9.2\times 10^{-12}\, \eta\,\, \frac{n_a\,
v_{s7}^3}{2}\,\, \theta^2 \,\, \enf.
\end{equation}
The efficiency of X-ray bremsstrahlung emission $\eta$ is
relatively low due to the Coulomb losses of fast particles,
providing  powerful optical and IR diffuse emission from the
interclump medium. Having in mind that $\eta < 10^{-5}$ in the
ambient medium [of solar abundance] for the energy band below
100 keV, one may conclude that $n_a \cdot v_{s7}^3 \cdot \theta^2 >
10^5$. That is a rather stringent requirement implying a high
efficiency of electron acceleration by a relatively slow shock
with $v_{s7} \gsim$ 1 and $n_a > 100 \cmc$ for $\theta \sim$ 10.
A more detailed confrontation of the observational data and the
model predictions, including GeV  \gr emission, will be described elsewhere.

An alternative interpretation of the observed hard X-ray sources
could be a fast massive ballistically moving ejecta fragment
interacting with a cloud. The fragment could drive a shock of
velocity about 1,000 $\kms$ in the inter-clump medium providing a
powerful source of X-ray emission both in continuum and lines
(Bykov 2003). The fragments being enriched with metals (of an
average nuclear charge $\langle Z\rangle$) have higher radiation
efficiency, since $\eta \propto\, \langle Z\rangle$. The energetic
problem of the bremsstrahlung emission models $\langle Z\rangle
\cdot n_a \cdot v_{s7}^3 \cdot \theta^2 > 10^5$ is alleviated. The
line signature around 6.2~keV required to fit the {\it RXTE PCA}
data could be naturally explained in that case as a Fe K line
complex predicted for the ejecta fragment model, though a
contribution from hidden accreting sources (AGNs or X-ray
binaries) cannot be excluded.

A powerful wind of the early type O9V star HD 193322 located just
7\arcmin\ from the NW bright clump (see Fig.~\ref{fig:A_R_I})
could also be considered as a candidate source for the observed
emission. The star is the central object of the open
cluster Collinder 419 located at an estimated distance 1.4 kpc
(e.g. McKibben et al., 1998)
which is close to the \gc distance estimations. The trigonometric parallax
of HD 193322 provided by the {\it Hipparcos} catalogue is
(2.1 $\pm$ 0.61) mas marginally consistent with the distance.
Interaction of the wind of HD 193322 with G78.2+2.1
would be a plausible source of
particle acceleration and hard X-ray emission.


A deeper exposure of the field with {\it ISGRI}\, is required to
detect [or to place meaningful upper limits] on the hard emission
from the NW, NE, SE regions up to
120 keV. The presence of a few clumps of hard emission at a level
above 10$^{-11} \enf$ at the borders of \gc would support the
hypothesis of SNR origin of the nonthermal clumps with important
implications for particle acceleration mechanisms in SNRs.

\begin{acknowledgements}

The present work is partly based on observations with INTEGRAL, an
ESA project with instruments and a science data centre funded by
ESA member states (especially the PI countries: Denmark, France,
Germany, Italy, Switzerland, Spain), Czech Republic and Poland,
and with the participation of Russia and the USA. This research
has made use of data obtained from the High Energy Astrophysics
Science Archive Research Center (HEASARC), provided by NASA's
Goddard Space Flight Center. We are grateful to F.Mavromatakis who
generously provided us with optical data.

The work was partially supported by RFBR grants 03-02-17433,
04-02-16595, 04-02-16042, RAS program, MK 2642.2003.02 and by the European Space
Agency. R.A.C. was supported by NASA grant NNG04GA41G. G.R. and J.P.S.
acknowledge support through the Belgian FNRS and the INTEGRAL PRODEX project.
Support from the International Space Science Institute (Bern) through the
international teams program is gratefully acknowledged.

\end{acknowledgements}

{\bf References} \\


Buckley, J.H., Akerlof, C.W., Carter-Lewis, D.A. et al. 1998,
A\&A, 329, 639

Bykov, A.M., Chevalier, R.A., Ellison, D.C. \& Uvarov, Yu.A. 2000,
ApJ, 538, 203

Bykov, A.M. 2003, A\&A, 410 ,L5

Chevalier, R.A. 1999,  ApJ,  511, 798

Esposito, J.A., Hunter, S.D., Kanbach, G., et al. 1996, ApJ, 461,
820


Gaisser, T., Protheroe, R., \& Stanev, T. 1998, ApJ, 492, 219

Goldwurm, A., David, P., Foschini, L. et al., 2003,  A\&A, 411,
L223

Higgs, L., Landecker, T., \& Roger, R. 1977, AJ, 82, 718

Landecker, T., Roger, R., \& Higgs, L. 1980, A\&AS, 39, 133

Lebrun, F., Leray, J.P., Lavocat, P.,  et al., 2003, A\&A, 411,
L141

Lozinskaya, T.A. 1977, Sov. Astron. Lett.,3,163.

Lozinskaya, T.A., Pravdikova, V.V., \& Finoguenov, A.V. 2000, Astron.
Lett., 26, 77

Lubinski, P., 2004, astro-ph/0405460.

Mavromatakis, F. 2003, A\&A, 408,  237

McKibben, W.P., Bagnuolo, W. G., Jr., Gies, D. R. et al.  1998, PASP, 110, 900

Pannuti, T., Allen, G., Houck, J. et al. 2003, ApJ, 593, 377

Sturner, S.J., \& Dermer, C.D.  1995, A\&A, 293,  L17

Sturner, S.J., Skibo, J.G., Dermer, C.D., \& Mattox, J.R. 1997,
ApJ, 490, 619

Swanenburg, B.N., Bennett, K., Bignami, G. F. et al. 1981, ApJ, 243, L69

Uchiyama, Y., Takahashi, T., Aharonian, F.A., \& Mattox, J.R. 2002,
ApJ, 571, 866

Winkler, C., Courvoisier, T., Di Cocco, G.,  et al. 2003, A\&A,
411, L1

Zhang, X., Zheng, Y., Landecker, T. L., \& Higgs, L. A.   1997, A\&A, 324,  641


\clearpage

\end{document}